\documentclass[aps,prb,preprint,showpacs,superscriptaddress,floatfix]{revtex4}
\usepackage{graphics}
\usepackage{epsfig}
\usepackage{amsmath}
\usepackage{amssymb}
\usepackage{calc}

\begin{document}

\title{Bound states in two-dimensional shielded potentials}
%\\
%Numerical evidence for the $\int{dr r V(r)}>0$ case}

\author{I. Nagy}
\affiliation{Department of Theoretical Physics, Technical University of Budapest,\\ 
H-1521 Budapest, Hungary}
\affiliation{Donostia International Physics Center DIPC, P. Manuel de Lardizabal 4,\\ 
20018 San Sebasti\'an, Spain}

\author{M.J. Puska}
\affiliation{Laboratory of Physics, Helsinki University of Technology,
P.O. Box 1100, FIN-02015 HUT, Finland}
\affiliation{Donostia International Physics Center DIPC, P. Manuel de Lardizabal 4,\\ 
20018 San Sebasti\'an, Spain}

\author{N. Zabala}
\affiliation{Elektrika eta Elektronika Saila, Zientzia eta Teknologia Fakultatea,\\
UPV-EHU 644 P.K., 48080 Bilbao, Spain}
\affiliation{Donostia International Physics Center DIPC, P. Manuel de Lardizabal 4,\\ 
20018 San Sebasti\'an, Spain}
\affiliation{Centro Mixto CSIC-UPV/EHU, P. Manuel de Lardizabal 4,\\ 
20018 San Sebasti\'an, Spain}

\date{\today}

\begin{abstract}
We study numerically the existence and character of bound states 
for positive and negative point charges shielded by the response
of a two-dimensional homogeneous electron gas. The problem is 
related to many physical situations and has recently arisen
in experiments for impurities on metal surfaces with
Schockley surface states. Mathematical theorems ascertain
a bound state for two-dimensional circularly symmetric potentials 
$V(r)$ with $\int_{0}^{\infty}{dr r V(r)} \leq 0$. We find that a
shielded potential with $\int_{0}^{\infty}{dr r V(r)} > 0$
may also sustain a bound state. Moreover, on the same footing 
we study the electron-electron
interactions in the two-dimensional electron gas finding a bound 
state with an energy minimum for a certain electron gas density.

\end{abstract}


\maketitle

\section{Introduction and motivation}

The properties of the two-dimensional (2D) electron gas and in particular the
phenomena induced by isolated or clustered impurities embedded in it have attracted
a large volume of experimental and theoretical research. The interest stems from the
increasingly more ideal realisations of the 2D electron gas, for example, at interfaces of
semiconductor heterostructures, at semiconductor surface inversion layers, at (noble)
metal surfaces as Schockley surface states or as quantum well states in metallic
overlayers on insulators or on other metals. At the same time the development of
different photoelectron and scanning tunneling spectroscopies has enabled an
increasingly more accurate characterization of these systems \cite{Hufner95,Echenique04}.

Many experiments deal directly or indirectly with the existence of bound electron states
in systems interacting with the 2D electron gas. The impurity-induced electron localization
has been studied in the Schockley surface state systems \cite{Olsson04,Limot05,Liu06}.
The quantum diffusion of hydrogen on metal surfaces reflects the coupling of the hydrogen
with the metallic (possibly 2D) electron gas \cite{Lauhon00,Kondo84}. Finally, a bound state
between two electrons in a 2D electron gas has been proposed as an alternative pairing
mechanism to the phonon coupling for high-temperature superconductivity \cite{Randeria89,Ghazali95}.
Remarkably, it was explicitly pointed out \cite{Randeria89} that the many-body ground state of a
dilute gas of fermions is unstable to pairing if and only if a two-body bound state exists.

The background for the above-mentioned physics is laid by basic mathematical theorems.
First, in two dimensions, for any everywhere attractive circularly symmetric potential,
$V(r)$, there exists always a bound state no matter how weak the potential is. 
For the case of a shallow potential valley characterized by the 
$\int_{0}^{\infty}{dr r V(r)}<0$ condition, an explicit expression for the binding 
energy, $E_b$, was derived in the textbook by Landau \cite{Landau58}.  
It turns out that the binding energy depends exponentially on the inverse
of a negative constant (given by the condition) 
in the investigated {\it weak}-coupling limit.
The theorem of Simon \cite{Simon76} extends the conditions for a bound state
to the $\int_{0}^{\infty}{dr r V(r)}=0$ case, i.e., to the not everywhere 
nonpositive or not everywhere nonnegative {\it suitable} potentials. 

In this work we present a numerical analysis for the existence of a 2D bound state
in potentials fulfilling the  $\int{dr r V(r)}={0}$ condition. 
The physically motivated effective potentials, $V_{eff}(r)={\Lambda}V(r)$,  
will refer to the perfectly shielded fields of 
embedded attractive or repulsive unit-charges in a 2D electron gas, and
${\Lambda}$ plays the role of a convenient coupling constant to detailed numerics.
In the repulsive case the screening is constrained by the fact that the maximum
of the surrounding hole density is the uniform electron gas density.
Beside the cases based on the mentioned standard condition, the overscreening 
and underscreening of unit-charges (with ${\Lambda}=1$) shall be investigated as well.
The corresponding potentials could mimic the shielding-dynamics in, for example, 
standing wave generation \cite{Limot05} on surfaces in the presence of impurities.
Slightly surprisingly we find a bound state also for the overscreened attractive and
underscreened repulsive potentials. In order to treat also the electron-electron
effective interaction we use in the Schr\" odinger equation the reduced mass ${\mu}=1/2$.
The energy of the ensuing bound electron pair has a minimum at a certain electron gas density.

The rest of the paper is organized as follows. In the next section, Sec. II, we shall deduce
our physically motivated model potentials to numerical, 2D bound-state calculations.
The results obtained are presented on illustrative figures, by considering relevant
parameter-ranges in coupling, screening-tuning, and density of the electron gas.
Finally, Sec. III is devoted to a short summary.
Hartree atomic units, ${\hbar}=e^2=m_e=1$, will be used in the equations 
and discussion below.

\section{Models and results}

\subsection{Shielded potentials}

We begin with few mathematical expressions. The 2D Fourier-Hankel (F-H) 
transformation of a $F(r)$ function is

\begin{equation}
 F(q)\, =\, 2{\pi}\, \int_{0}^{\infty}\, dr\, r\, J_{0}(rq)\,F(r),
\end{equation}
where $J_{0}(x)$ is the zeroth-order Bessel function. The inverse 
F-H transform has the form

\begin{equation}
 F(r)\, =\, \frac{1}{2\pi}\, \int_{0}^{\infty}\, dq\, q\, J_{0}(qr)\,F(q).
\end{equation}
These equations will be used below, in model-potential constructions.

The field of an embedded charged particle is shielded in the 2D 
electron gas. Thus instead of the bare Coulomb form, $v_{c}(q)=2\pi/q$, 
one can write in momentum space 

\begin{equation}
V(q)\, =\, {\pm}v_{c}(q)\, [1\, -\, {\Delta}n(q)],
\end{equation}
for the shielded field around unit-charges of different signs. Here 
${\Delta}n(q)$ is the screening density in momentum-space.  
With a unit-norm ${\Delta}n(r)$ one obtains, via the above Eqs.(1) and (3), 
the $V(q\rightarrow{0})\propto{q}$ limiting behaviour {\it if}
the real-space density decays faster than $r^{-4}$. 
This case corresponds to the $\int{dr r V(r)}=0$ condition of Simon's theorem. 

Notice, that the well-known \cite{Stern67} quasiclassical, Thomas-Fermi ($TF$) 
approximation for impurity screening in a 2D electron gas
results in a {\it monotonic} potential

\begin{equation}
V_{TF}(r)\, =\, {\pm}\, \int_{0}^{\infty}\, dq \frac{q}{q + 2}\, J_0(qr)\, =\,
{\pm}\, \int_{0}^{\infty}\, dx \frac{xe^{-2}}{(x^2 + r^2)^{3/2}}.
\end{equation}
%
%in which $s=2$ due to the special character of the density of states in the 
Negative and positive signs in Eq.(4) refer to embedded attractive 
and repulsive unit charges, respectively. At large distances 
($r\rightarrow{\infty}$) 
the $V_{TF}(r)$ potential falls of as ${\pm}1/(4r^3)$, 
and one gets the $\int_{0}^{\infty}{dr r V_{TF}(r)}={\pm}{(1/2)}$ condition.
Therefore, this type of potential with negative sign belongs 
to the class analyzed by Landau \cite{Landau58}.

We shall characterize the ${\Delta}n(r)$ density {\it directly}
by the properly normalized hydrogenic $(H)$
and gaussian $(G)$ forms, $({\alpha}^2/{2\pi})\, exp(-{\alpha}r)$ and 
$({\beta}^2/{\pi})\, exp(-{\beta}^2r^2)$, respectively.
The corresponding Fourier-Hankel transforms are ${\alpha}^{3}/({\alpha}^2 + q^2)^{3/2}$
and $exp[-q^{2}/(2\beta)^{2}]$, respectively. 
Using the former in Eq.(3) and applying Eq.(2), we get 

\begin{equation}
V_{H}(r)\, =\, {\pm}\, \frac{1}{r}\left(1 - 2u^2\, [I_{0}(u)K_{1}(u) - I_{1}(u)K_{0}(u)]\right),
\end{equation}
in which $u={\alpha}r/2$ for shorthand. The gaussian-form results in

\begin{equation}
V_{G}(r)\, =\, {\pm}\, \frac{1}{r}\left(1 - \sqrt{2{\pi}z}\, I_{0}(z)\, e^{-z}\right),
\end{equation}
where $z={\beta}^2r^2/2$. In the above equations $I_{i}(x)$ and $K_{i}(x)$ 
are modified Bessel functions. 

We stress that $V(q=0)=0$, i.e., $\int_{0}^{\infty}{dr r V(r)}=0$ in these 
models. For long distances these shielded potentials decay as
${\mp}{3/(2{\alpha}^{2}r^{3}})$, and
${\mp}{1/(4{\beta}^{2}r^{3}})$, respectively.
The attractive potentials are exhibited in Fig. 1, by fixing ${\alpha}=4$ 
(solid curve)
and ${\beta}=2\sqrt{2}$ (dashed curve). The ${\alpha}=4$ value could correspond
to a strictly atomistic, $1s$-like in 2D, electron density. 
For a detailed comparison, two other potentials are also plotted. That is, the
bare Coulomb one, $V_{c}(r)=(-1/r)$, and the conventional Thomas-Fermi 
form, $V_{TF}(r)$ of Eq.(4).

\begin{figure}
\centerline{\resizebox{9.5cm}{!}{\includegraphics{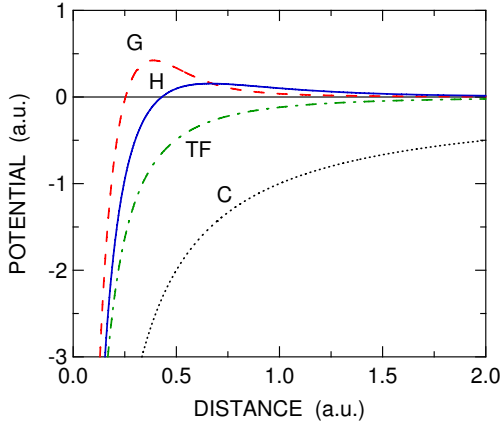}}}
\caption{\label{fig1} (Color online)
The shielded attractive potentials of Eq.(5) (solid blue curve) and Eq.(6) 
(dashed red curve) with ${\alpha}=4$ and ${\beta}=2\sqrt{2}$, respectively.
The dotted black curve refers to a bare Coulomb potential, $-1/r$.
The dash-dotted green curve is devoted to the attractive Thomas-Fermi 
form of Eq.(4).}
\end{figure}

In order to go beyond the $\int_{0}^{\infty}{dr r V(r)}=0$ condition, 
one may multiply the {\it shielding} parts of Eqs.(5) and (6) by a 
${\lambda}$ variable [See Eqs. (\ref{attrV}) and (\ref{repV}) below]. 
In such a way we can investigate the overscreened 
(${\lambda}>1$) and underscreened (${\lambda}<1$) cases, 
for which $\int_{0}^{\infty}{dr r V(r,{\lambda})}\ne{0}$.
Moreover, as discussed above the effective potentials have
a multiplicative coupling constant $\Lambda$,
[$V_{eff}(r)={\Lambda}V(r,\lambda)$].  
In our 2D numerics ${\Lambda}$ and ${\lambda}$ will serve 
as convenient parameters.

\subsection{Numerical solution of the 2D Schr\"odinger equation}

The bound-state energy levels ($E_{b}$) and wave functions [${\psi}(r)$] 
satisfy the 2D Schr\" odinger equation

\begin{equation}
\left[-\, \frac{1}{2\mu}\, {\nabla}^{2}\, +\, V_{eff}(r)\, -E_{b}\right]\, 
{\psi}({\bf \rm r})\, =\, 0,
\label{schr}
\end{equation}
where ${\mu}$ is the reduced mass; it is unity for the impurity-electron case.

In circular symmetry the wave function separates as

\begin{equation}
{\psi}({\bf \rm r})\, =\, \frac{e^{i m \phi}}{\sqrt{2\pi}}
R_{mn}(r),
\end{equation}
where $m=0,\pm{1},\pm{2}....$ is the azimuthal quantum number and 
$n=1,2,3,....$ is the radial quantum number related to the number 
of radial nodes $(n-1)$ of the radial wave function $R_{mn}(r)$. 
In this work we are only interested in the values $m=0$ and $n=1$. 
Further, by making the substitution $U_{mn}(r)=r^{1/2}R_{mn}(r)$ we obtain the
differential equation

\begin{equation}
\frac{d^{2}U_{mn}(r)}{dr^{2}} +
\left(2\mu[E_{b} - V_{eff}(r)] -
\frac{(m^{2}-1/4)}{r^{2}}\right)
U_{mn}(r)=0.
\end{equation}

This is the same form as the radial equation studied in spherically 
symmetric problems. We solve the equation on an exponentially expanding 
radial mesh, $r(j)=r_{min}\exp((j-1)\Delta x)$ with $j$=1 \dots $N$. 
With a given guess for the eigenvalue $E_{b}$ the function $U_{mn}(r)$ is
integrated outwards from origin and inwards from a large radius by 
starting with its asymptotic expansions. At a matching point close to 
the classical turning point the logarithmic derivatives of the outward 
and inward integrated solutions are required to coincide by adjusting 
the eigenvalue $E_{b}$. The parameters of the radial mesh, $r_{min}$, 
${\Delta x}$, and $N$ are varied until the numerical convergence of the
eigenvalue is obtained.

\subsection{Attractive electron-ion interaction}

According to our numerical calculations, the attractive $V_{TF}(r)$ of 
Eq.(4) gives $E_{b}^{(TF)}=-0.2853$, while the $V_{H}(r)$ of
Eq.(5) results in $E_{b}^{(H)}=-0.0296$ values for ${\alpha}$=4, 
in Hartree units. There is an about order-of-magnitude reduction 
in binding due to an atomistic screening, beyond quasiclassics. 

In order to study the so-called not everywhere nonpositive case 
(physically: the shielded positive charge) in more detail
we will use, without loss of generality, ${\alpha}=4$ in
Eq.(5) and modify it as
\begin{equation}
V_{eff}^H (r)\, =\, -\, {\Lambda}\, \frac{1}{r}\left(1 - 8{\lambda}r^2\, 
[I_{0}(2r)K_{1}(2r) - I_{1}(2r)K_{0}(2r)]\right).
\label{attrV}
\end{equation}
In Fig. 2 we have plotted the $E_{b}^{(H)}(\lambda)$ energies obtained with 
${\Lambda}$=1 and ${\lambda}\in[0.97,1.03]$. 
The ${\lambda}$-tuning refers to small under- and overscreening.
The figure clearly shows the sensitivity of binding on the details of 
shielding. For ${\lambda}>1$ one has $\int_{0}^{\infty}{dr r V_{eff}(r)}>{0}$, 
with Eq.(\ref{attrV}).

\begin{figure}
\centerline{\resizebox{9.5cm}{!}{\includegraphics{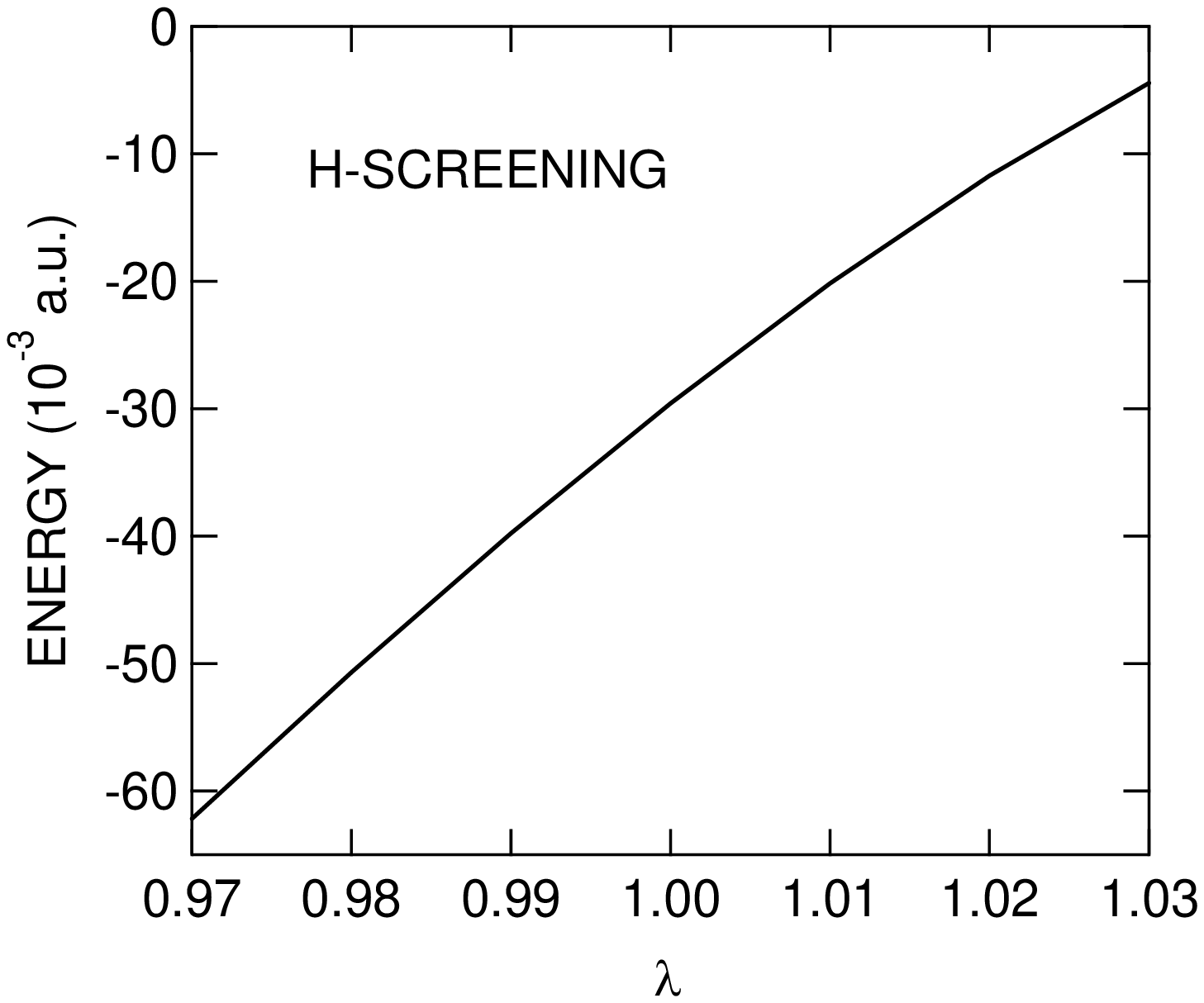}}}
\caption{\label{fig2} 
Binding energy, $E_{b}^{(H)}(\lambda)$, based on Eq.(7) with 
Eq.(\ref{attrV}) for ${\Lambda}$=1 and
${\lambda}\in[0.97,1.03]$.}
\end{figure}

Additional information on the ${\Lambda}$-dependence of $E_{b}^{(H)}(\Lambda)$ 
for Eq.(\ref{attrV}) and ${\lambda}$=1, are given in Fig. 3 with 
${\Lambda}\in[1.25,0.75]$.
In this case $\int_{0}^{\infty}{dr r V_{eff}(r)}={0}$ for any (finite) ${\Lambda}$,
as we pointed out earlier.
In harmony with Simon's theorem \cite{Simon76},             
$|E_{b}(\Lambda)|\le{exp(-c{\Lambda}^{-2})}$,
we get for our case [$E_{b}^{(H)}(\Lambda)$] an about $c=5.292$ value for 
the suitable constant in the investigated coupling-parameter range.
The macroscopic dielectric constant of a real medium can result in reduced, 
${\Lambda}<1$ values to effective interactions.

In this attractive impurity case we have shown a remarkable sensitivity of
theoretical bound-state characteristics to shielding conditions. This is in
accord with important spectroscopic information obtained \cite{Limot05}
by scanning tunneling spectroscopy for different adatoms in generated standing
wave patterns. The observed peak-shift and amplitude-decrease in differential
tunneling conductance, as adatoms are approached to a step on the
surface, signal the experimental sensitivity.

We note that our detailed numerical analysis is based on an effective
Schr\" odinger equation. Further attempts are needed therefore to consider the
many-body aspects of the localization problem in more detail. For example, the
{\it proper} description of the width of an adatom-induced 
weakly bound state and its influence on scattering characteristics 
are important questions to future studies.

\subsection{Repulsive electron-electron interaction}

The so-called not everywhere nonnegative case (physically: the shielded 
negative charge) will be treated similarly via Eq.(5) of positive sign, 
but with the constraint
\begin{equation}
{\lambda}{\alpha}^2/{2\pi}\equiv{n_0}=1/({\pi}r_s^2)
\end{equation}
in order to model bounded, complete depletion of the
2D electron gas density at $r=0$. 
Here $r_s$ is the density-parameter of the 2D gas 
with density $n_0$. 
The same constraint is used to the corresponding (positive sign) reparametrized
Eq.(6) with 
\begin{equation}
{\lambda}{\beta}^2/{\pi}\equiv{n_0}=1/({\pi}r_s^2) 
\end{equation}
as
\begin{equation}
V_{eff}^G(r)\, =\, +\, {\Lambda}\, \frac{1}{r}
\left(1 - {\lambda}\sqrt{2{\pi}z}\, I_{0}(z)\, e^{-z}\right),
\label{repV}
\end{equation}
where, according to the above discussion, $z=(r/r_s)^{2}/(2\lambda)$ now.
Remember, that at ${\lambda}=1$ one has $\int_{0}^{\infty}{dr r V_{eff}(r)}=0$,
independently of the value of a finite ${\Lambda}$. 

As we use the depletion-constraint-based densities to model effective 
electron-electron interactions, it is important to give an additional physical 
argument on their proper behaviour. The interaction energy (${\varepsilon}$) 
of a repulsive point charge with the surrounding, normalized (${\lambda}=1$) 
shielding hole can be characterized by the classical equation

\begin{equation}
{\varepsilon}\, =\, -\, \frac{1}{2}\, {2\pi} \int_{0}^{\infty} dr\, r\, \frac{1}{r}\, 
{\Delta}n(r).
\label{holeint}
\end{equation}
This equation results in the ${\varepsilon}_{H}=-1/(\sqrt{2}r_s)$, 
and ${\varepsilon}_{G}=-\sqrt{\pi}/(2r_s)$
expressions for the exponential and gaussian screening, respectively. 
These energies are {\it between} the values based on
the exchange-only, ${\varepsilon}_{x}=-4\sqrt{2}/(3{\pi}r_s)$, and 
Wigner-monatom, \cite{Wigner34,Nagy99}
${\varepsilon}_{W}=-1/r_s$, limiting approximations. 
The latter corresponds to the ${\Delta}n(r\le{r_s})=n_0$ extremum model
for the hole-density, while the former to the Pauli-hole 
\cite{Seidl99} of an ideal 2D system.

\begin{figure}
\centerline{\resizebox{9.5cm}{!}{\includegraphics{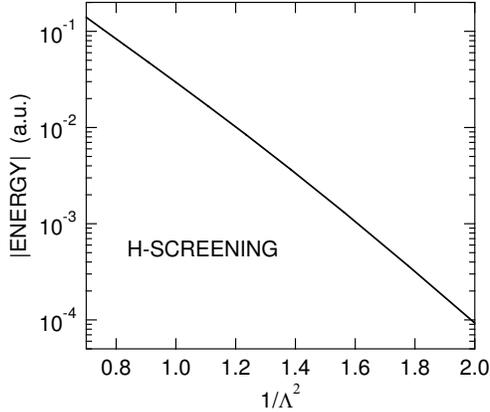}}}
\caption{\label{fig3} 
 Binding energy, $E_{b}^{(H)}(\Lambda)$, based on Eq.(7) with Eq.(10) for ${\lambda}$=1 and
${\Lambda}\in[1.25,0.75]$.}
\end{figure}

The two repulsive potentials, discussed above in details, are
used in Eq.(\ref{schr}) with the reduced mass ${\mu}=1/2$,
and the following numerical results are obtained.
The binding energies 
$E_{b}(r_s)$ are plotted in Fig. 4,
as a function of the density parameter $r_s$. Computed values are connected by solid
(the case of exponential shielding) and dashed (the case of gaussian model) curves. The
other parameters behind these results are fixed as ${\Lambda}={\lambda}=1$.

\begin{figure}
\centerline{\resizebox{9.5cm}{!}{\includegraphics{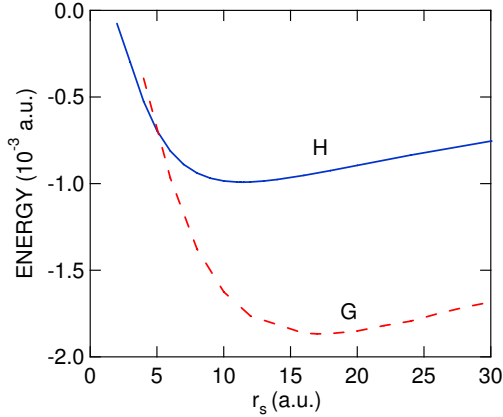}}}
\caption{\label{fig4} (Color online)
Binding energies, $E_{b}(r_s)$, in the shielded field of a negative charge. Data are
based on Eq.(\ref{schr}) with ${\mu}=1/2$, and ${\Lambda}={\lambda}=1$ in the applied two
effective interactions. The solid blue and dashed red curves refer 
to  hydrogenic [Eq. (5)] and gaussian [Eq. (6)] models, respectively.
See the text for further details.}
\end{figure}

Remarkably, there are $r_s$ parameter values at which the binding energies 
are optimal, i.e., they have extremal values. By increasing or decreasing 
the density of the electron gas, the bindings become weaker. 
The  extremal values are at about ${r_s}=11.4$ for the hydrogen-like
model, and at ${r_s}=17.2$ for the gaussian model. The binding energies are 
in the $10^{-3}$ range in atomic units, for the dilute system. 
The low carrier density and thus a small Fermi energy, 
${\varepsilon}_{F}=1/r_s^2$, are {\it important} characteristics of 
cuprate superconductors \cite{Uemura91}. 
In these materials the ratio of the critical temperature ($T_c$) and 
the Fermi energy is in the range of $10^{-1}$. Furthermore, there 
is a saturation and supression of $T_c$ with increasing carrier density; 
for further detail we refer to Fig. 3 of Ref.[16].

The observed extremal character is related, physically, to our 
pseudononlinear construction of effective shielding of repulsive charges. 
Namely, the constraint via the bounded depletion hole at $r=0$ fixes 
a reasonable scaling in the potentials. In standard linear-response 
theory \cite{Ghazali95} the hole-density at $r=0$ can become
higher than the host density $n_0$. In such an attempt one can get a monotonic
dependence of $E_b(r_s)$ on $r_s$ in the resulting shielded fields.

Further information is given in Figs. 5 and 6, which show the 2D radial densities,
$2{\pi}r|{\psi}(r)|^{2}$, computed with {\it bound-state} wave functions at the extremal
Wigner-Seitz parameters. The corresponding potentials are also plotted.
As expected, the square-integrable wave functions are localized at about the
potential minima. In the investigated equal-mass case this extension can represent
a certain coherence-lenght; somewhat surprisingly it is only twice of the extremal density parameter.

\begin{figure}
\centerline{\resizebox{9.5cm}{!}{\includegraphics{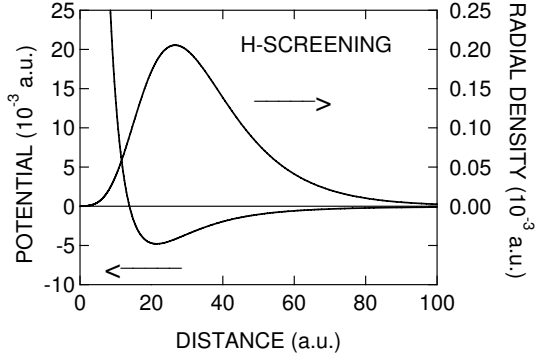}}}
\caption{\label{fig5} 
The radial density $2{\pi}r|{\psi}(r)|^{2}$ and the potential based on
hydrogenic screening of a negative unit-charge. These are computed at
the $r_s=11.4$ value of the density parameter.}         
\end{figure}

\begin{figure}
\centerline{\resizebox{9.5cm}{!}{\includegraphics{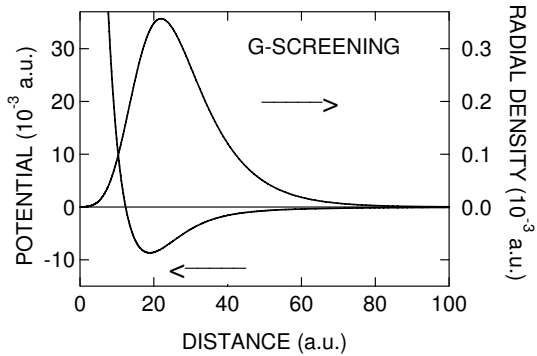}}}
\caption{\label{fig6} 
The radial density $2{\pi}r|{\psi}(r)|^{2}$ and the potential based on
gaussian screening of a negative unit-charge. These are computed at
the $r_s=17.2$ value of the density parameter.}    
\end{figure}

As we observed in Fig. 3 for the attractive case, the binding energy depends on the coupling, 
${\Lambda}$. An illustration on this fact for the repulsive case is given in Fig. 7. 
The gaussian model of Eq.(\ref{repV}) is used with ${\lambda}=1$, ${\mu}=1/2$, and 
$r_s=17.2$. Results for ${\Lambda}<1$ are plotted. We can approximate our data
by a quadratic expression, i.e., $|E_{b}^{(G)}({\Lambda})|\sim{{\Lambda}^2}$.
Rescaling of ${\Lambda}=1$ by a macroscopic dielectric constant could result in
a notable reduction of the above-mentioned (see, Fig. 4, also) binding energies.

\begin{figure}
\centerline{\resizebox{9.5cm}{!}{\includegraphics{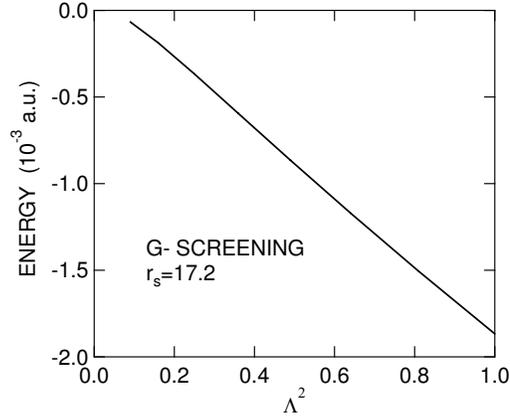}}}
\caption{\label{fig7} 
The coupling-constant (${\Lambda}$) dependence of the binding energy for the repulsive case. 
The results are based on the gaussian model of Eq.(\ref{repV}) with ${\mu}=1/2$ in Eq.(\ref{schr}).
The Wigner-Seitz parameter is $r_s=17.2$ and ${\lambda}=1$.}             
\end{figure}

We finish our representation, by discussing the question of undershielding the repulsive charge. 
Fig. 8 is devoted to this problem. We have used Eq.(\ref{repV}) with 
${\Lambda}=1$ and ${\lambda}<1$ in
the Schr\" odinger equation, Eq.(\ref{schr}) with ${\mu}=1/2$, at $r_s=17.2$. One can observe (see, Fig. 2, 
for the attractive case) essential reductions of energies for the 
$\int_{0}^{\infty}{dr r V(r)}>0$ unconventional condition.

\begin{figure}
\centerline{\resizebox{9.5cm}{!}{\includegraphics{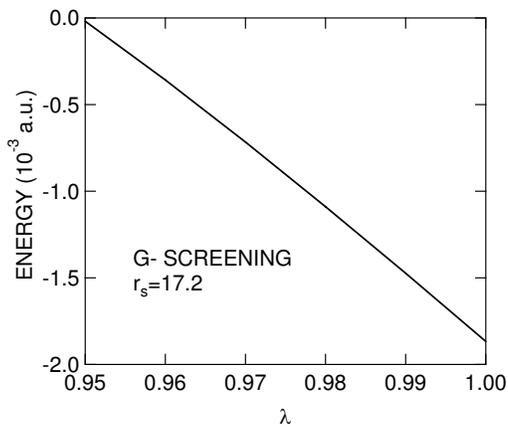}}}
\caption{\label{fig8} 
The shielding-constant (${\lambda}$) dependence of the binding energy for the repulsive case. 
The results are based on the gaussian model of Eq.(\ref{repV}) with 
${\mu}=1/2$ in Eq.(\ref{schr}).
The Wigner-Seitz parameter is $r_s=17.2$ and ${\Lambda}=1$.}             
\end{figure}

The last results of Figs. 7 and 8, together with Figs. 4-6,
signal a nontrivial sensitivity of the magnitude of the binding energy
on the concrete physical situation. Fortunately, there is a physical 
limitation. It should be clear from the discussion at Eq.(\ref{holeint}), 
and as our 
illustrative figures indeed show, that a more localized real-space character 
of the bounded (and normalized) depletion hole around a repulsive 
unit-charge, results in an effective potential with
a repulsive part of a shorter range. Clearly, the limitation is given by the 
Wigner model, in which one has the ${\Delta}n(r\le{r_s})=n_0$ extremum for 
the hole.

\section{Summary}
In this work we have investigated the problem of bound states in two-dimensional
shielded potentials. Effective potentials, based on direct approximation for the
screening charge densities around attractive and repulsive unit charges, are employed.
Particularly, the effective electron-electron interaction is modelled via a 
properly constrained depletion hole. 
In the detailed numerical analysis performed, we found that in both of the
basically attractive, and basically repulsive potentials bound states appear
under the $\int_{0}^{\infty}{dr r V(r)}={0}$ standard, and may appear under the 
$\int_{0}^{\infty}{dr r V(r)}>{0}$ unconventional conditions. 
In the repulsive case an extremal character of the binding energy, as a function 
of the density of the host 2D electron gas, is established with our physical models
for effective electron-electron interactions.

\section{Acknowledgments}
I.N. and M.J.P. are thankful for the warm hospitality at DIPC, where this study 
was completed. They gratefully acknowledge useful discussions with P.M. Echenique. 
The work of I.N. has been supported partly by the Hungarian OTKA 
(Grant Nos. T046868 and T049571), that of M.J.P. by the Academy of 
Finland, and that of N.Z. by the University of Basque Country
(9/UPV00206.215-13639/2001) and the Basque Unibertsitate eta
Ikerketa Saila, the MCyT (FIS2004-06490-C03-00) and the EU Network of Excellence 
NANOQUANTA (NMP4-CT-2004-500198).


\newpage

\begin{thebibliography}{}
\bibitem{Hufner95}
S. H\"ufner, {\it Photoelectron
Spectroscopy - Principles and Applications}, vol. 82 of Springer Series 
in Solid-State Science (Springer, Berlin, 1995).
%
\bibitem{Echenique04}
P.M. Echenique, R. Berndt, E.V. Chulkov, Th. Fauster, A. Goldman, and U. H\" ofer,
Surf. Sci. Reports {\bf 52}, 219 (2004).
%
\bibitem{Olsson04}
F.E. Olsson, M. Persson, A.G. Borisov, J.P. Gauyacq, J. Lagoute, and S. F\" olsch,
Phys. Rev. Letters {\bf 93}, 206803 (2004).
%
\bibitem{Limot05}
L. Limot, E. Pehlke, J. Kr\" oger, and R. Berndt,
Phys. Rev. Letters {\bf 94}, 036805 (2005).
%
\bibitem{Liu06}
C. Liu, I. Matsuda, R. Hobara, and S. Hasegawa,
Phys. Rev. Letters {\bf 96}, 036803 (2006).
%
\bibitem{Lauhon00}
L.J. Lauhon and W. Ho, Phys. Rev. Letters {\bf 85}, 4566 (2000).
%
\bibitem{Kondo84}
J. Kondo, Physica (Amsterdam) {\bf 125B}, 279 (1984).
%
\bibitem{Randeria89}
M. Randeria, J.M. Duan, and L.Y. Shieh, Phys. Rev. Letters {\bf 62}, 981 (1989).
%
\bibitem{Ghazali95}
A. Ghazali and A. Gold, Phys. Rev. B {\bf 52}, 16634 (1995).
%
\bibitem{Landau58}
L.D. Landau and E.M. Lifshitz, {\it Quantum Mechanics}
(Addison-Wesley, Massachusetts, 1958).
%
\bibitem{Simon76}
B. Simon, Ann. Phys. (New York) {\bf 97}, 279 (1976).
%
\bibitem{Stern67}
F. Stern and W.E. Howard, Phys. Rev. {\bf 163}, 816 (1967).
%
\bibitem{Wigner34}
E. Wigner, Phys. Rev. {\bf 46}, 1002 (1934).
%
\bibitem{Nagy99}
I. Nagy, Phys. Rev. B {\bf 60}, 4404 (1999).
%
\bibitem{Seidl99}
M. Seidl, J.P. Perdew, and M. Levy, Phys. Rev. A {\bf 59}, 51 (1999).
%
\bibitem{Uemura91}
Y.J. Uemura {\it et al.}, Phys. Rev. Letters {\bf 66}, 2665 (1991).
%
%\bibitem{Zaremba05}
%E. Zaremba, I. Nagy, and P.M. Echenique,
%Phys. Rev. B {\bf 71}, 125323 (2005).
%

\end{thebibliography}
\end{document}